\documentclass[twocolumn,showpacs,preprintnumbers]{revtex4}%
\usepackage{amssymb}
\usepackage{amsmath}
\usepackage{graphicx}
\usepackage{dcolumn}
\usepackage{bm}
\usepackage{amsfonts}%
\providecommand{\U}[1]{\protect\rule{.1in}{.1in}}
\providecommand{\U}[1]{\protect\rule{.1in}{.1in}}
\providecommand{\U}[1]{\protect\rule{.1in}{.1in}}
\begin{document}
\title{Synchronization in an Optomechanical Cavity}
\author{Keren Shlomi}
\author{D.Yuvaraj}
\author{Ilya Baskin}
\author{Oren Suchoi}
\author{Roni Winik}
\author{Eyal Buks}
\affiliation{Department of Electrical Engineering, Technion, Haifa 32000 Israel}
\date{\today }

\begin{abstract}
We study self-excited oscillations (SEO) in an on-fiber optomechanical cavity.
Synchronization is observed when the optical power that is injected into the
cavity is periodically modulated. A theoretical analysis based on the
Fokker-Planck equation evaluates the expected phase space distribution (PSD)
of the self-oscillating mechanical resonator. A tomography technique is
employed for extracting PSD from the measured reflected optical power.
Time-resolved state tomography measurements are performed to study phase
diffusion and phase locking of the SEO. The detuning region inside which
synchronization occurs is experimentally determined and the results are
compared with the theoretical prediction.

\end{abstract}
\maketitle





\section{Introduction}

Optomechanical cavities \cite{Braginsky&Manukin_67,
Hane_179,Gigan_67,Metzger_1002,Kippenberg_1172,Favero_104101,Marquardt2009}
are widely employed for various sensing \cite{Rugar1989, Arcizet2006a,
Forstner2012,Weig2013} and photonics applications
\cite{Lyshevski&Lyshevski_03,Stokes_et_al_90, Hossein_Zadeh_276,Wu_et_al_06,
MattEichenfield2007,Bahl2011,Flowers-Jacobs2012}. Moreover, such systems may
allow experimental study of the crossover between classical to quantum realms
\cite{Thompson_72, Meystre2013,Kimble_et_al_01, Carmon_et_al_05,
Arcizet_et_al_06, Gigan_67, Jayich_et_al_08, Schliesser_et_al_08,
Genes_et_al_08, Teufel_et_al_10,Poot_273}. The effect of radiation pressure
typically governs the optomechanical coupling (i.e. the coupling between the
electromagnetic cavity and the mechanical resonator that serves as a movable
mirror) when the finesse of the optical cavity is sufficiently high
\cite{Kippenberg_et_al_05,Rokhsari2005,
Arcizet2006,Gigan_et_al_06,Cooling_Kleckner06, Kippenberg_1172}, whereas,
bolometric effects can contribute to the optomechanical coupling when optical
absorption by the vibrating mirror is significant \cite{Metzger_1002,
Jourdan_et_al_08,Marino&Marin2011PRE, Metzger_133903, Restrepo_860,
Liberato_et_al_10,Marquardt_103901, Paternostro_et_al_06,Yuvaraj_430}.
Generally, bolometric effects are dominant in systems comprising of relatively
large mirrors in which the thermal relaxation rate is comparable to the
mechanical resonance frequency \cite{Aubin_et_al_04, Marquardt_103901,
Paternostro_et_al_06, Liberato_et_al_10_PRA}. These systems
\cite{Metzger_133903, Metzger_1002, Aubin_et_al_04,
Jourdan_et_al_08,Zaitsev_046605,Zaitsev_1589} exhibit many intriguing
phenomena such as mode cooling and self-excited oscillations (SEO)
\cite{Hane_179,Kim_1454225,Aubin_1018,Carmon_223902,Marquardt_103901,Corbitt_021802,Carmon_123901,Metzger_133903}%
. It has been recently demonstrated that optomechanical cavities can be
fabricated on the tip of an optical fiber \cite{iannuzzi06, Ma2010,
iannuzzi10, iannuzzi11, Jung2011,
Butsch2012,Albri2013,Shkarin_1306_0613,Baskin_1210_7327,Yuvaraj_210403}. These
micron-scale devices, which can be optically actuated \cite{Iannuzzi2013}, can
be used for sensing physical parameters that affect the mechanical properties
of the suspended mirror (e.g. absorbed mass, heating by external radiation,
acceleration, etc.).%

\begin{figure}
[ptb]
\begin{center}
\includegraphics[
trim=0.000000in 0.000000in -0.154059in 0.000000in,
height=2.514in,
width=3.2482in
]%
{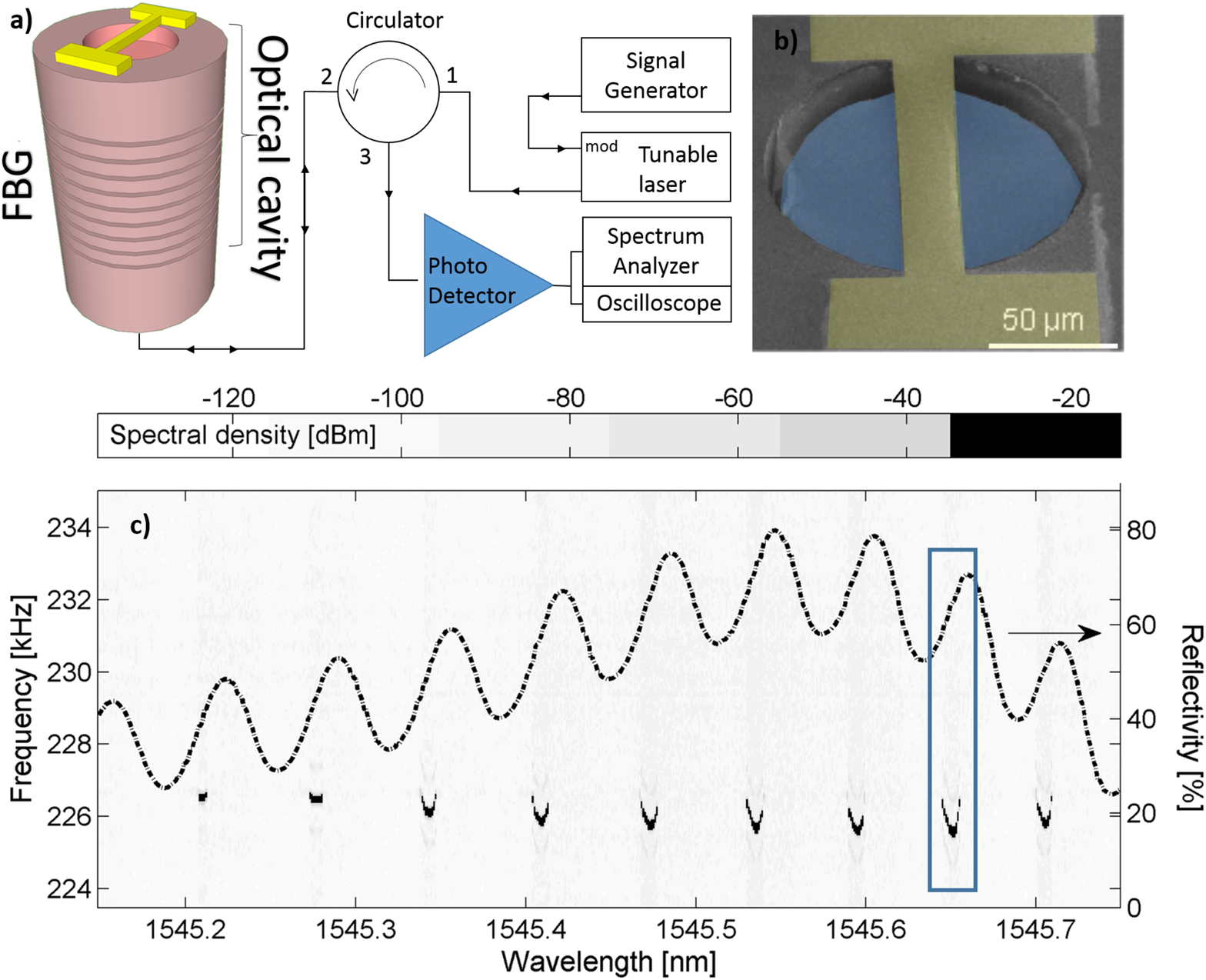}%
\caption{Experimental setup. (a) A schematic drawing of the sample and the
experimental set-up. An on-fiber optomechanical cavity is excited by a tunable
laser with modulated power. The reflected light intensity is measured and
analyzed. (b) Electron micrograph of a suspended micromechanical mirror (false
color code: blue-silica fiber, yellow - gold mirror, gray - zirconia ferrule),
the view is tilted by $52^{0}$. (c) Spectral decomposition of the reflected
light power $P_{\mathrm{R}}$ vs. excitation wavelengths $\lambda_{\mathrm{L}}%
$. The SEO, visible as sharp peaks (black regions on colormap) in the
reflected power spectrum, are obtained at optical excitation wavelengths
corresponding to positive slopes of the sample's reflectivity (shown by a
dotted curve). The cavity resonance used in the synchronization experiments is
denoted by a rectangle.}%
\label{Fig setup}%
\end{center}
\end{figure}

In the present study we optically induce SEO \cite{Rugar1989, Arcizet2006a,
Forstner2012,Weig2013} by injecting a monochromatic laser light into an
on-fiber optomechanical cavity, which is formed between a fiber Bragg grating
(FBG) mirror, serving as a static reflector, and a vibrating mirror, which is
fabricated on the tip of a single mode optical fiber. These optically-induced
SEO are attributed to the bolometric optomechanical coupling between the
optical mode and the mechanical resonator \cite{Zaitsev_046605,Zaitsev_1589}.
We find that the phase of the SEO can be synchronized by periodically
modulating the laser power that is injected into the cavity.

Synchronization \cite{Pikovsky_3}, one of the most fundamental phenomena in
nature, has been observed since 1673 \cite{Hugenii_1} in many different setups
and applications \cite{Blekhman_1,Blekhman_2,Aulova_504,Schafer_857, Peskin_1,
Hop_1}. Synchronization in self-oscillating systems
\cite{Rosenblum_401,Osipov_1,
Pikovsky_2291,Afraimovich_1,Kuznetsov_221,Landa_1,Fradkov_1} can be the result
of interaction between systems \cite{Pecora_2374, Pecora_821, de_R7359,
Fujisaka_32,Landa_414, Warminski_677,Lemonde_053602}, external noise
\cite{Pikovsky_576, Balanov_L113, Czolczynski_937, Balanov_041105, Zhang_411,
Rosenblum_1804, Pikovsky_219,Yang_1753} or other outside sources, periodic
\cite{Koronovskii_847, Nikitin_171,Min_202} or non-periodic
\cite{Nakabayashi_163,Rosenblum_264102}. Synchronization can also be activated
by delayed feedback \cite{Janson_010601, Balanov_1, Scholl_281, Hamdi_1}.

Here we employ the technique of state tomography
\cite{Vogel_2847,Yuvaraj_210403} in order to experimentally measure the phase
space distribution (PSD) of the mechanical element near the threshold of SEO.
Time resolved tomography \cite{Suchoi_1408_2331} is employed in order to
monitor the process of phase diffusion. Furthermore, we study the response of
the system to periodic modulation of the laser power. We witness phase locking
at certain regions of modulation amplitude and modulation frequency, for which
the SEO are synchronized with the external modulation
\cite{Anishchenko_117,Pandey_3,Paciorek_1723,Adler_351,Jensen_1637,DosSantos_1147}%
. The experimental results are compared with theoretical predictions that are
obtained by solving the Fokker-Planck equation that governs the dynamics of
the system.

\section{Experimental Setup}

The optomechanical cavity shown in Fig. \ref{Fig setup} was fabricated on the
flat polished tip of a single mode fused silica optical fiber with outer
diameter of $126%
\operatorname{\mu m}%
$ (Corning SMF-28 operating at wavelength band around $1550%
\operatorname{nm}%
$) held in a zirconia ferrule. A $10%
\operatorname{nm}%
$-thick chromium layer and a $200%
\operatorname{nm}%
$ gold layer were successively deposited by thermal evaporation. The bilayer
was directly patterned by a focused ion beam to the desired mirror shape ($20%
\operatorname{\mu m}%
$-wide doubly clamped beam). Finally, the mirror was released by etching
approximately $12%
\operatorname{\mu m}%
$ of the underlying silica in 7\% HF acid ($90%
\operatorname{min}%
$ etch time at room temperature). The beam remained supported by the zirconia
ferrule, which is resistant to HF. The precise alignment between the
micro-mechanical mirror and fiber core that was achieved in the fabrication
process allowed robust and simple operation without any need for
post-fabrication positioning.

The static mirror of the optomechanical cavity was provided by a fiber Bragg
grating (FBG) mirror (made using a standard phase mask technique
\cite{Anderson_566}, grating period of $0.527%
\operatorname{\mu m}%
$ and length $\approx8%
\operatorname{mm}%
$) with the reflectivity band of $0.4%
\operatorname{nm}%
$ full width at half maximum (FWHM) centered at $1550%
\operatorname{nm}%
$. The length of the optical cavity was $l\approx10%
\operatorname{mm}%
$, providing a free spectral range of $\Delta\lambda=\lambda_{0}%
^{2}/2n_{\mathrm{eff}}l\approx80$ pm (where $n_{\mathrm{eff}}$ $=1.468$ is the
effective refraction index for SMF-28). The cavity length was chosen so that
at least five cavity resonance wavelengths would be located within the range
of the FBG reflectivity band. Despite the high FBG reflectivity ($\approx
90\%$), the resulting cavity finesse was low (about 2) due to the high cavity
losses (see Ref. \cite{Zaitsev_046605} for detailed discussion of the cavity
reflectivity spectrum). The most plausible source of losses is the light
scattering on the rough etched fiber tip surface (micron size protuberances
were observed below the suspended beam), giving rise to radiation loss.

Monochromatic light was injected into the fiber bearing the cavity on its tip
from a laser source with an adjustable output wavelength ($\lambda
_{\mathrm{L}}$, tunable in the range of $1527.6-1565.5%
\operatorname{nm}%
$) and power level $P_{\mathrm{L}}$. The laser was connected through an
optical circulator, that allowed the measurement of the reflected light
intensity $P_{\mathrm{R}}$ by a fast responding photodetector. The detected
signal was analyzed by an oscilloscope and a spectrum analyzer (see the
schematics in Fig. \ref{Fig setup}). The experiments were performed in vacuum
(at residual pressure below $0.01%
\operatorname{Pa}%
$) at a base temperature of $77%
\operatorname{K}%
$.

\section{Fokker-Planck Equation}

The micromechanical mirror in the optical cavity is treated as a mechanical
resonator with a single degree of freedom $x$ having mass $m$ and linear
damping rate $\gamma_{0}$ (when it is decoupled from the optical cavity). It
is assumed that the angular resonance frequency of the mechanical resonator
depends on the temperature $T$ of the suspended mirror. For small deviation of
$T$ from the base temperature $T_{0}$ (i.e. the temperature of the supporting
substrate) it is taken to be given by $\omega_{0}-\beta T_{\mathrm{R}}$, where
$T_{\mathrm{R}}=T-T_{0}$ and where $\beta$ is a constant. Furthermore, to
model the effect of thermal deformation \cite{Metzger_133903} it is assumed
that a temperature dependent force given by $m\theta T_{\mathrm{R}}$, where
$\theta$ is a constant, acts on the mechanical resonator \cite{Yuvaraj_430}.
When noise is disregarded, the equation of motion governing the dynamics of
the mechanical resonator is taken to be given by%
\begin{equation}
\frac{\mathrm{d}^{2}x}{\mathrm{d}t^{2}}+2\gamma_{0}\frac{\mathrm{d}%
x}{\mathrm{d}t}+\left(  \omega_{0}-\beta T_{\mathrm{R}}\right)  ^{2}x=\theta
T_{\mathrm{R}}\;. \label{x eom}%
\end{equation}

The intra-cavity optical power incident on the suspended mirror is denoted by
$P_{\mathrm{L}}I\left(  x\right)  $, where $P_{\mathrm{L}}$ is the injected
laser power, and the function $I\left(  x\right)  $ depends on the mechanical
displacement $x$ [see Eq. (\ref{I(x)}) below]. The time evolution of the
relative temperature $T_{\mathrm{R}}$ is governed by the thermal balance
equation%
\begin{equation}
\frac{\mathrm{d}T_{\mathrm{R}}}{\mathrm{d}t}=Q-\kappa T_{\mathrm{R}}\;,
\label{T_R eom}%
\end{equation}
where $Q=\eta P_{\mathrm{L}}I\left(  x\right)  $ is proportional to the
heating power, $\eta$ is the heating coefficient due to optical absorption and
$\kappa$ is the thermal decay rate.

The function $I\left(  x\right)  $ depends on the properties of the optical
cavity that is formed between the suspended mechanical mirror and the on-fiber
static reflector. The finesse of the optical cavity is limited by loss
mechanisms that give rise to optical energy leaking out of the cavity. The
main escape routes are through the on-fiber static reflector, through
absorption by the metallic mirror, and through radiation. The corresponding
transmission probabilities are respectively denoted by $T_{\mathrm{B}}$,
$T_{\mathrm{A}}$ and $T_{\mathrm{R}}$. In terms of these parameters, the
function $I\left(  x\right)  $ is given by \cite{Zaitsev_046605}%
\begin{equation}
I\left(  x\right)  =\frac{\beta_{\mathrm{F}}\left(  1-\frac{\beta_{-}^{2}%
}{\beta_{+}^{2}}\right)  \beta_{+}^{2}}{1-\cos\frac{4\pi x_{\mathrm{D}}%
}{\lambda}+\beta_{+}^{2}}\;, \label{I(x)}%
\end{equation}
where $x_{\mathrm{D}}=x-x_{\mathrm{R}}$ is the displacement of the mirror
relative to a point $x_{\mathrm{R}}$, at which the energy stored in the
optical cavity in steady state obtains a local maximum, $\beta_{\pm}%
^{2}=\left(  T_{\mathrm{B}}\pm T_{\mathrm{A}}\pm T_{\mathrm{R}}\right)
^{2}/8$ and where $\beta_{\mathrm{F}}$ is the cavity finesse. The reflection
probability $R_{\mathrm{C}}=P_{\mathrm{R}}/P_{\mathrm{L}}$ is given in steady
state by \cite{Yurke_5054,Zaitsev_046605} $R_{\mathrm{C}}=1-I\left(  x\right)
/\beta_{\mathrm{F}}$. The function $I\left(  x\right)  $ can be expanded as
$I\left(  x\right)  =I_{0}+I_{0}^{\prime}x+\left(  1/2\right)  I_{0}%
^{\prime\prime}x^{2}+O\left(  x^{3}\right)  $, where a prime denotes
differentiation with respect to the displacement $x$.

Consider the case where the laser power $P_{\mathrm{L}}$ is periodically
modulated in time according to%
\begin{equation}
P_{\mathrm{L}}=P_{0}+P_{1}\cos\left(  \omega_{\mathrm{p}}t\right)  \;,
\end{equation}
where $P_{0}$, $P_{1}$ and $\omega_{\mathrm{p}}$ are constants. When both
$P_{1}$ and $I-I_{0}$ are sufficiently small, the following approximation can
be employed%
\begin{equation}
Q=\eta P_{\mathrm{L}}I\simeq\eta P_{0}I+\eta P_{1}I_{0}\cos\left(
\omega_{\mathrm{p}}t\right)  \;. \label{Q heating}%
\end{equation}
For the case where $\kappa t\gg1,$ the solution of Eq. (\ref{T_R eom}) can be
expressed as%
\begin{equation}
T_{\mathrm{R}}=T_{\mathrm{R}0}+T_{\mathrm{R}1}\;, \label{T_R}%
\end{equation}
where $T_{\mathrm{R}0}$ is a solution of Eq. (\ref{T_R eom}) for the case
where the laser power is taken to be the constant $P_{0}$, and where
$T_{\mathrm{R}1}$, which is given by%
\begin{equation}
T_{\mathrm{R}1}=\frac{\eta P_{1}I_{0}\cos\left(  \omega_{\mathrm{p}}%
t-\phi_{\mathrm{p}}\right)  }{\sqrt{\kappa^{2}+\omega_{\mathrm{p}}^{2}}}\;,
\label{T_R1}%
\end{equation}
where $\tan\phi_{\mathrm{p}}=\omega_{\mathrm{p}}/\kappa$, represents the
temperature variation due to the power modulation with a fixed displacement.

Substituting the expansion (\ref{T_R}) into Eq. (\ref{x eom}), neglecting
terms of second order in $\beta$ and disregarding the phase $\phi_{\mathrm{p}%
}$ (i.e. shifting time by $\phi_{\mathrm{p}}/\omega_{\mathrm{p}}$) yield%
\begin{align}
&  \frac{\mathrm{d}^{2}x}{\mathrm{d}t^{2}}+2\gamma_{0}\frac{\mathrm{d}%
x}{\mathrm{d}t}+\omega_{\mathrm{m}}^{2}\left[  1+\zeta\cos\left(
\omega_{\mathrm{p}}t\right)  \right]  x\nonumber\\
&  =f_{\text{\textrm{th}}}+f_{\text{\textrm{e}}}\cos\left(  \omega
_{\mathrm{p}}t\right)  \;,\nonumber\\
&  \label{x eom mod}%
\end{align}
where $\omega_{\mathrm{m}}^{2}=\omega_{0}^{2}-2\omega_{0}\beta T_{\mathrm{R}%
0}$ is the temperature dependent angular resonance frequency, $\zeta
=-2\beta\eta P_{1}I_{0}/\omega_{0}\sqrt{\kappa^{2}+\omega_{\mathrm{p}}^{2}}$
is the amplitude of parametric excitation due to laser power modulation [see
Eq. (\ref{T_R1})], $f_{\text{\textrm{th}}}=\theta T_{\mathrm{R}0}$ is the
thermal force, and $f_{\text{\textrm{e}}}=\theta\eta P_{1}I_{0}/\sqrt
{\kappa^{2}+\omega_{\mathrm{p}}^{2}}$ is the force amplitude due to laser
power modulation [see Eq. (\ref{T_R1})]. Furthermore, as was mentioned above,
the temperature $T_{\mathrm{R}0}$ is assumed to satisfy [see Eq.
(\ref{T_R eom})]%
\begin{equation}
\frac{\mathrm{d}T_{\mathrm{R}0}}{\mathrm{d}t}=\eta P_{0}I\left(  x\right)
-\kappa T_{\mathrm{R}0}\;. \label{T_RO eom}%
\end{equation}

As can be seen from Eq. (\ref{x eom mod}), modulating the laser power gives
rise to two contributions, one representing parametric excitation with
amplitude $\zeta$ originating from the temperature dependence of the resonance
frequency, and another representing direct forcing with amplitude
$f_{\text{\textrm{e}}}$ originating from the thermal force term. Both these
terms can be treated using the rotating wave approximation (RWA) only when the
angular frequency $\omega_{\mathrm{p}}$ is chosen to be close to particular
values. Two such values are considered below, $\omega_{0}$ and $2\omega_{0}$.
When $\omega_{\mathrm{p}}\simeq\omega_{0}$ the effect of the direct forcing
term is expected to dominate, whereas when $\omega_{\mathrm{p}}\simeq
2\omega_{0}$ the effect of the parametric term is expected to dominate. These
two cases can be simultaneously treated by assuming that in Eq.
(\ref{x eom mod}) $\omega_{\mathrm{p}}=\omega_{0}+\omega_{\mathrm{d}}$ in the
direct forcing term and $\omega_{\mathrm{p}}=2\left(  \omega_{0}%
+\omega_{\mathrm{d}}\right)  $ in the parametric term, where $\omega
_{\mathrm{d}}\ll\omega_{0}$ is the detuning.

The displacement $x\left(  t\right)  $ can be expressed in terms of the
complex amplitude $A$ as $x\left(  t\right)  =x_{0}+2\operatorname{Re}\left(
Ae^{i\omega_{\mathrm{p}}t}\right)  $, where $x_{0}$, which is given by
$x_{0}=\eta\theta P_{0}I_{0}/\kappa\omega_{0}^{2}$, is the optically-induced
static displacement. Assuming that $A$ is small and it is slowly varying on
the time scale of $\omega_{0}^{-1}$ and applying the RWA yield a first order
evolution equation for the complex amplitude $A=A_{x}+iA_{y}$, where both
$A_{x}$ and $A_{y}$ are real \cite{Zaitsev_1589}, which can be written in a
vector form as%
\begin{equation}
\mathbf{\dot{A}}+\mathbf{\Phi}=\mathbf{\xi}_{\mathrm{R}}\;, \label{A dot V}%
\end{equation}
where $\mathbf{A}=\left(  A_{x},A_{y}\right)  $, the vector $\mathbf{\Phi
}=\left(  \Phi_{x},\Phi_{y}\right)  $, where both $\Phi_{x}$ and $\Phi_{y}$
are real, is given by
\begin{equation}
\mathbf{\Phi}=\nabla\mathcal{H}+\omega_{\mathrm{d}}\left(  -A_{y}%
,A_{x}\right)  \;, \label{Phi}%
\end{equation}
the scalar function$\mathcal{\ H}$ is given by \cite{Zaitsev_859}%
\begin{align}
\mathcal{H}  &  =\frac{\Gamma_{0}\left(  A_{x}^{2}+A_{y}^{2}\right)  }%
{2}+\frac{\Gamma_{2}\left(  A_{x}^{2}+A_{y}^{2}\right)  ^{2}}{4}\nonumber\\
&  +\frac{\omega_{0}\zeta}{4}A_{x}A_{y}-\frac{f_{\text{\textrm{e}}}A_{x}%
}{\omega_{0}}\;,\nonumber\\
&  \label{H P}%
\end{align}
$\Gamma_{0}=\gamma_{0}+\eta\theta P_{\mathrm{L}}I_{0}^{\prime}/2\omega_{0}%
^{2}$ is the effective rate of linear damping, $\Gamma_{2}=\gamma_{2}%
+\eta\beta P_{\mathrm{L}}I_{0}^{\prime\prime}/4\omega_{0}$ is the effective
nonlinear quadratic damping rate and $\gamma_{2}$ is the intrinsic mechanical
contribution to $\Gamma_{2}$. The noise term $\mathbf{\xi}_{\mathrm{R}%
}=\left(  \xi_{\mathrm{R}x},\xi_{\mathrm{R}y}\right)  $, where both
$\xi_{\mathrm{R}x}$ and $\xi_{\mathrm{R}y}$ are real, satisfies $\left\langle
\xi_{\mathrm{R}x}\left(  t\right)  \xi_{\mathrm{R}x}\left(  t^{\prime}\right)
\right\rangle =\left\langle \xi_{\mathrm{R}y}\left(  t\right)  \xi
_{\mathrm{R}y}\left(  t^{\prime}\right)  \right\rangle =2\tau\delta\left(
t-t^{\prime}\right)  $ and $\left\langle \xi_{\mathrm{R}x}\left(  t\right)
\xi_{\mathrm{R}y}\left(  t^{\prime}\right)  \right\rangle =0$, where
$\tau=\gamma_{0}k_{\mathrm{B}}T_{\mathrm{eff}}/4m\omega_{0}^{2}$,
$k_{\mathrm{B}}$ is the Boltzmann's constant and $T_{\mathrm{eff}}$ is the
effective noise temperature.

In the absence of laser modulation, i.e. when $P_{1}=0$, the equation of
motion (\ref{A dot V}) describes a van der Pol oscillator \cite{Pandey_3}.
Consider the case where $\Gamma_{2}>0$, for which a supercritical Hopf
bifurcation occurs when the linear damping coefficient $\Gamma_{0}$ vanishes.
Above threshold, i.e. when $\Gamma_{0}$ becomes negative, the amplitude
$A_{r}=\left\vert A\right\vert =\sqrt{A_{x}^{2}+A_{y}^{2}}$ of SEO is given by
$A_{r0}=\sqrt{-\Gamma_{0}/\Gamma_{2}}$.

Consider the case of vanishing detuning, i.e. the case where $\omega
_{\mathrm{d}}=0$, for which $\mathbf{\Phi}=\nabla\mathcal{H}$. For this case,
the Langevin equation (\ref{A dot V}) for the complex amplitude $A$ yields the
corresponding Fokker-Planck equation for the PSD $\mathcal{P}\left(
A_{x},A_{y}\right)  $, which can be written as
\cite{Hempstead_350,Risken_Fokker-Planck}%
\begin{equation}
\frac{\partial\mathcal{P}}{\partial t}-\nabla\cdot\left(  \mathcal{P}%
\nabla\mathcal{H}\right)  -\tau\nabla\cdot\left(  \nabla\mathcal{P}\right)
=0\;. \label{Fokker-Planck P}%
\end{equation}

\section{Synchronization}

The steady state solution $\mathcal{P}_{0}$ of (\ref{Fokker-Planck P}) is
given by \cite{Risken_Fokker-Planck}%
\begin{equation}
\mathcal{P}_{0}=\frac{1}{Z}\exp\left(  -\frac{\mathcal{H}}{\tau}\right)  \;,
\label{P_0}%
\end{equation}
where $Z$ is a normalization constant (partition function). We experimentally
investigate the effect of laser power modulation for the above discussed two
cases, i.e. $\omega_{\mathrm{p}}=\omega_{0}$ and $\omega_{\mathrm{p}}%
=2\omega_{0}$, and compare the results to the theoretical prediction given by
Eq. (\ref{P_0}) [recall that Eq. (\ref{P_0}) is valid only when the detuning
vanishes, i.e. when $\omega_{\mathrm{d}}=0$]. For both cases, the PSD is
extracted from the measured off reflected cavity power using the technique of
state tomography \cite{Vogel_2847,Yuvaraj_210403}.

\begin{figure}
[ptb]
\begin{center}
\includegraphics[
height=4.1277in,
width=3.5008in
]%
{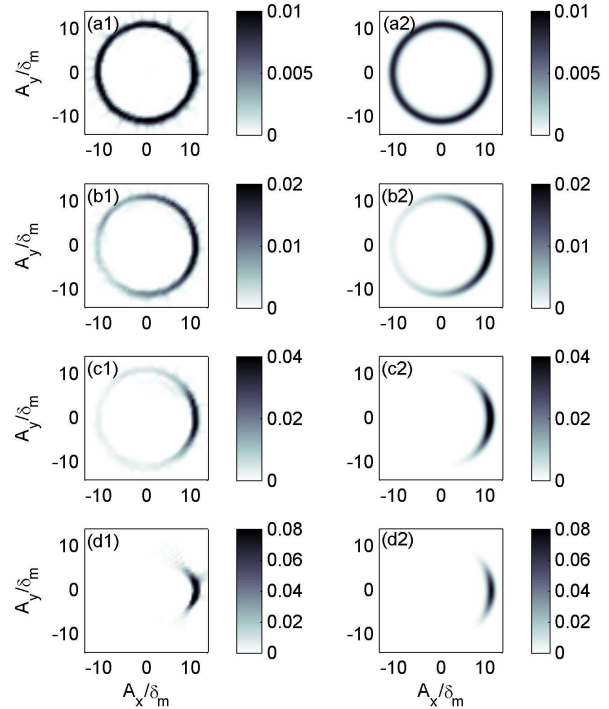}%
\caption{The PSD as a function of modulation amplitude at resonance
$\omega_{\mathrm{p}}=\omega_{0}$. The panels on the left exhibit the measured
PSD whereas the panels on the right exhibit the calculated PSD obtained from
Eq. (\ref{P_0}). The modulation amplitude in the panels labeled as a, b, c and
d is $P_{1}/P_{0}=0.3\times10^{-3}$, $0.8\times10^{-3}$, $3.3\times10^{-3}$
and $6.7\times10^{-3}$, respectively. The following device parameters have
been employed in order to calculate the PSD according to Eq. (\ref{P_0}):
$m=1.1\times10^{-12}\operatorname{kg}$, $\omega_{0}=2\pi\times
225\operatorname{kHz}$ and $\left(  I_{0}/I_{0}^{\prime}\delta_{\mathrm{m}%
}\right)  \left(  1+\kappa^{2}/\omega_{\mathrm{p}}^{2}\right)  ^{-1/2}=25$.}%
\label{Fig P_0 f_e}%
\end{center}
\end{figure}

\begin{figure}
[ptb]
\begin{center}
\includegraphics[
height=3.1704in,
width=3.5163in
]%
{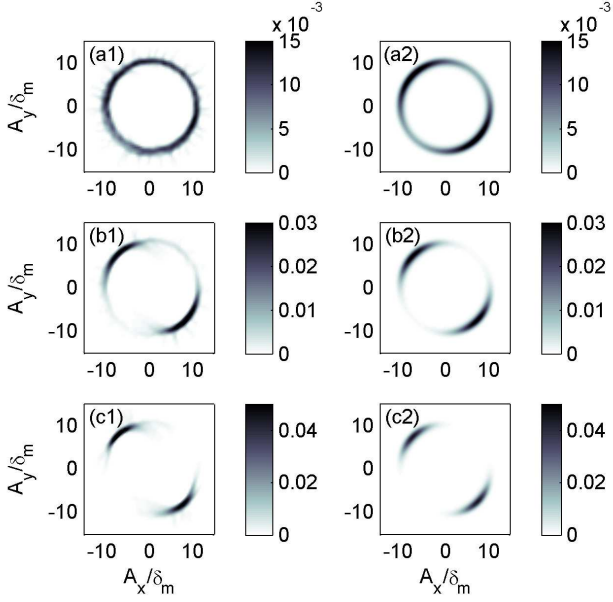}%
\caption{The PSD as a function of modulation amplitude at angular frequency
$\omega_{\mathrm{p}}=2\omega_{0}$. The panels on the left exhibit the measured
PSD whereas the panels on the right exhibit the calculated PSD obtained from
Eq. (\ref{P_0}). The modulation amplitude in the panels labeled as a, b and c
is $P_{1}/P_{0}=0.67\times10^{-2}$, $2\times10^{-2}$ and $3.3\times10^{-2}$,
respectively. The parameter $\lambda_{\mathrm{L}}\beta\omega_{0}/\theta=4800$
together with the other parameters that are listed in the caption of Fig.
\ref{Fig P_0 f_e} have been employed in order to calculate the PSD according
to Eq. (\ref{P_0}).}%
\label{Fig P_0 zeta}%
\end{center}
\end{figure}

The results that are obtained with $\omega_{\mathrm{p}}=\omega_{0}$ are seen
in Fig. \ref{Fig P_0 f_e}. For this case, the laser wavelength is
$\lambda_{\mathrm{L}}=1545.641%
\operatorname{nm}%
$ and the average power is $P_{0}=12%
\operatorname{mW}%
$ (the data seen in Figs. \ref{Fig P_0 zeta}, \ref{Fig dephasing} and
\ref{Fig rephasing} was taken with the same values of $\lambda_{\mathrm{L}}$
and $P_{0}$). The panels on the left exhibit the measured PSD whereas the
panels on the right exhibit the calculated PSD obtained from Eq. (\ref{P_0}).
For both cases, the PSD is plotted as a function of the normalized coordinates
$A_{x}/\delta_{\mathrm{m}}$ and $A_{y}/\delta_{\mathrm{m}}$, where
$\delta_{\mathrm{m}}=\sqrt{2\tau/\gamma_{0}}$. The relative modulation
amplitude $P_{1}/P_{0}$ is increased from top to bottom (see figure caption
for the values). The ring-like shape of the PSD, which is seen in the top
panels, in which the relative modulation amplitude $P_{1}/P_{0}$ obtains its
lowest value, changes into a crescent-like shape as $P_{1}/P_{0}$ is
increased. While a PSD having a ring-like shape corresponds to SEO with a
random phase, synchronization gives rise to a PSD having a crescent-like
shape. The characteristic length of the crescent depends on both the
modulation amplitude and the noise intensity in the system. The device
parameters that have been employed in the theoretical calculation are listed
in the figure caption.%

The results that are obtained with $\omega_{\mathrm{p}}=2\omega_{0}$ are seen
in Fig. \ref{Fig P_0 zeta}. The relative modulation amplitude $P_{1}/P_{0}$ is
increased from top to bottom (see figure caption for the values). For this
case of modulation at $\omega_{\mathrm{p}}=2\omega_{0},$ synchronization gives
rise to two preferred values of the phase of SEO, which differ one from the
other by $\pi$, as can be seen from the double-crescent shape of both measured
and calculated PSD (see Fig. \ref{Fig P_0 zeta}).%

\section{Dephasing and Rephasing}

\begin{figure}
[ptb]
\begin{center}
\includegraphics[
height=3.1704in,
width=3.5163in
]%
{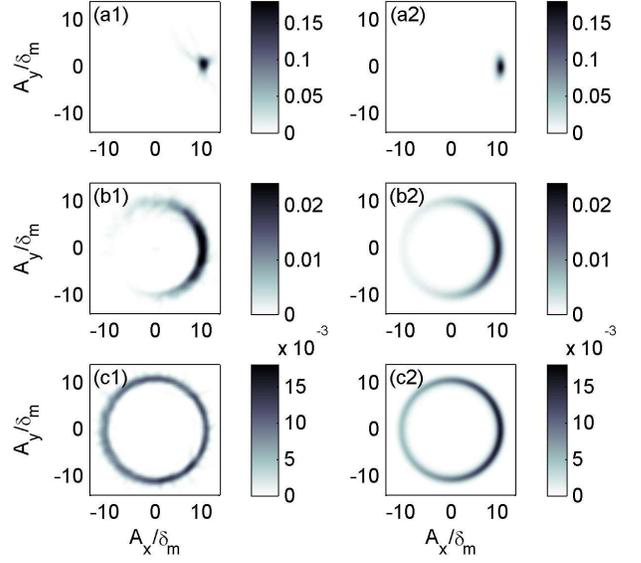}%
\caption{Dephasing of SEO. The panels on the left exhibit the measured PSD
whereas the panels on the right exhibit the calculated PSD obtained from
numerically integrating the Fokker-Planck equation (\ref{Fokker-Planck P}).
The normalized dwell time $\gamma_{0}t_{\mathrm{d}}$ in the panels labeled as
a, b and c is $\gamma_{0}t_{\mathrm{d}}=0.15$, $2.5$ and $75$, respectively.
The device parameters are the same as those given in the caption of Fig.
\ref{Fig P_0 f_e}.}%
\label{Fig dephasing}%
\end{center}
\end{figure}

The phase of SEO in steady state randomly drifts in time due to the effect of
external noise. In addition, noise gives rise to amplitude fluctuations around
the average value $A_{r0}$. To experimentally study these effects, SEO are
driven using the same parameters of laser power and wavelength as in Figs.
\ref{Fig P_0 f_e}\ and \ref{Fig P_0 zeta}. The off-reflected signal from the
optical cavity is recorded in two time windows separated by a dwell time
$t_{\mathrm{d}}$. While the data taken in the first time window is used to
determine the initial phase of SEO, the data taken in the second one is used
to extract PSD by state tomography \cite{Yuvaraj_210403} using the initial
phase as a reference phase. The results are seen in Fig. \ref{Fig dephasing}
for 3 different values of the dwell time $t_{\mathrm{d}}$ (given in the figure
caption). While the left panels show the measured PSDs, the panels on the
right exhibit the calculated PSDs obtained by numerically integrating the
Fokker-Planck equation (\ref{Fokker-Planck P}). The process of dephasing of
SEO is demonstrated by the transition from a PSD having a crescent-like shape
that is obtained for a relatively short dwell time $t_{\mathrm{d}}$ (see top
panels) to a PSD having a ring-like shape that is obtained for a relatively
long dwell time $t_{\mathrm{d}}$ (see bottom panels).%

\begin{figure}
[ptb]
\begin{center}
\includegraphics[
height=3.1704in,
width=3.5163in
]%
{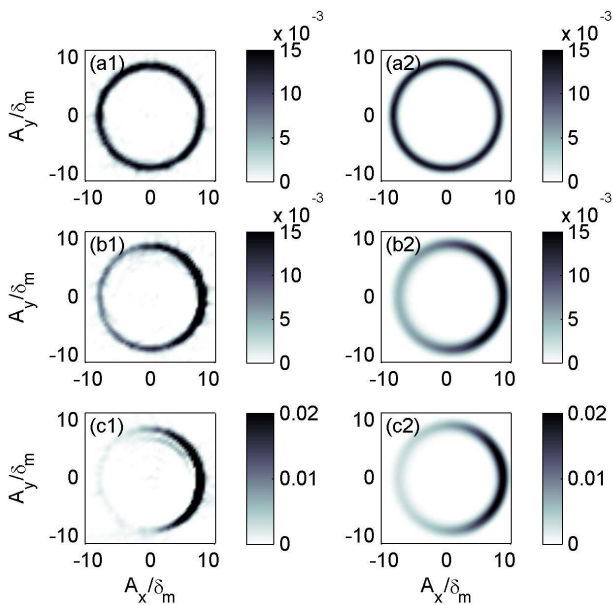}%
\caption{Rephasing of SEO. The relative amplitude of the modulation, which is
turned on at time $t_{0}=1\operatorname{s}$ after the first time window, is
$P_{1}/P_{0}=0.01$. The normalized dwell time $\gamma_{0}t_{\mathrm{d}}$ in
the panels labeled as a, b and c is $\gamma_{0}t_{\mathrm{d}}=0.05$, $0.95$
and $1.7$, respectively. The device parameters are the same as those given in
the caption of Fig. \ref{Fig P_0 f_e}. The panels on the left exhibit the
measured PSD whereas the panels on the right exhibit the calculated PSD
obtained from numerically integrating the Fokker-Planck equation
(\ref{Fokker-Planck P}).}%
\label{Fig rephasing}%
\end{center}
\end{figure}

The opposite process to dephasing, which is hereafter referred to as
rephasing, is demonstrated in Fig. \ref{Fig rephasing}. As was done in the
previous experiment, the off-reflected signal from the optical cavity is
recorded in two time windows separated by a dwell time, which is labeled for
the current case as $t_{0}+t_{\mathrm{d}}$. In addition, power modulation at
resonance (i.e. with $\omega_{\mathrm{p}}=\omega_{0}$) is turned on at time
$t_{0}=1%
\operatorname{s}%
$ after the first time window. The time $t_{0}$ is chosen to be much longer
than the dephasing time, and consequently the phase of SEO is fully randomized
at time $t_{0}$. While in Fig. \ref{Fig P_0 f_e} above, the case of
synchronization in steady state, i.e. in the limit of $t_{\mathrm{d}%
}\rightarrow\infty$, is demonstrated, in the current experiment the PSD is
measured for finite values of $t_{\mathrm{d}}$ in order the monitor in time
the process of rephasing. Contrary to the case of dephasing (see Fig.
\ref{Fig dephasing}), rephasing is demonstrated by the transition from a PSD
having a ring-like shape that is obtained for a relatively short dwell time
$t_{\mathrm{d}}$ (see top panels in Fig. \ref{Fig rephasing}) to a PSD having
a crescent-like shape that is obtained for a relatively long dwell time
$t_{\mathrm{d}}$ (see bottom panels).%

\section{Detuning Range of Phase Locking}

The region in the plane of modulation frequency $\omega_{\mathrm{p}}$ and
modulation amplitude $f_{\text{\textrm{e}}}$ in which synchronization occurs
can be determined by finding the fixed points of Eq. (\ref{A dot V}) and by
analyzing their stability \cite{Anishchenko_117,Pandey_3}. Consider the case
where $\omega_{\mathrm{p}}\simeq\omega_{0}$. For this case both the parametric
term and the noise term are disregarded, and thus $\mathbf{\Phi}=\left(
\Phi_{x},\Phi_{y}\right)  $ becomes [see Eq. (\ref{Phi})]%
\begin{align}
\Phi_{x}  &  =\left[  \Gamma_{0}+\Gamma_{2}\left(  A_{x}^{2}+A_{y}^{2}\right)
\right]  A_{x}-\omega_{\mathrm{d}}A_{y}-f_{\text{\textrm{e}}}\;,\\
\Phi_{y}  &  =\left[  \Gamma_{0}+\Gamma_{2}\left(  A_{x}^{2}+A_{y}^{2}\right)
\right]  A_{y}+\omega_{\mathrm{d}}A_{x}\;.
\end{align}
At a fixed point, i.e. when $\Phi_{x}=\Phi_{y}=0$, the following holds%
\begin{equation}
\mathcal{F}^{2}=\left[  \left(  1-\mathcal{A}^{2}\right)  ^{2}+\mathcal{D}%
^{2}\right]  \mathcal{A}^{2}\;, \label{F^2 C}%
\end{equation}
where $\mathcal{F}=f_{\text{\textrm{e}}}/A_{r0}\Gamma_{0}$ is the normalized
modulation amplitude, $\mathcal{A}=A_{r}/A_{r0}$ is the normalized radial
coordinate, $A_{r0}=\sqrt{-\Gamma_{0}/\Gamma_{2}}$ is the amplitude of SEO,
and $\mathcal{D}=\omega_{\mathrm{d}}/\Gamma_{0}$ is the normalized detuning.

The Jacobian matrix is given by%
\begin{equation}
J=\left(
\begin{array}
[c]{cc}%
\frac{\partial\Phi_{x}}{\partial A_{x}} & \frac{\partial\Phi_{x}}{\partial
A_{y}}\\
\frac{\partial\Phi_{y}}{\partial A_{x}} & \frac{\partial\Phi_{y}}{\partial
A_{y}}%
\end{array}
\right)  \;.
\end{equation}
The eigenvalues $\lambda_{\pm}$ of $J$ can be expressed in terms of the trace
$\operatorname{Tr}J=2\Gamma_{0}\left(  1-2\mathcal{A}^{2}\right)  $ and
determinant $\det J=\Gamma_{0}^{2}\left(  3\mathcal{A}^{4}-4\mathcal{A}%
^{2}+1+\mathcal{D}^{2}\right)  $ of $J$ as%
\begin{equation}
\lambda_{\pm}=\frac{\operatorname{Tr}J\pm\sqrt{\left(  \operatorname{Tr}%
J\right)  ^{2}-4\det J}}{2}\;.
\end{equation}

Hopf bifurcation occurs when $\operatorname{Tr}J=0$, i.e. when%
\begin{equation}
\mathcal{A}^{2}=\frac{1}{2}\;, \label{A^2=1/2}%
\end{equation}
and when $\det J>0$, i.e. when $\mathcal{A}^{2}<\mathcal{A}_{-}^{2}$ or
$\mathcal{A}^{2}>\mathcal{A}_{+}^{2}$, where%
\begin{equation}
\mathcal{A}_{\pm}^{2}=\frac{2}{3}\pm\frac{1}{3}\sqrt{1-3\mathcal{D}^{2}}\;.
\end{equation}
Hopf bifurcation is thus possible only when $\left\vert \mathcal{D}\right\vert
>0.5$ [see Eq. (\ref{A^2=1/2})]. Furthermore, combining Eqs. (\ref{F^2 C}) and
(\ref{A^2=1/2}) yields a relation between the modulation amplitude
$\mathcal{F}$ and the detuning $\mathcal{D}$ along the bifurcation line%
\begin{equation}
8\mathcal{F}^{2}=1+4\mathcal{D}^{2}\;. \label{F^2 Hopf}%
\end{equation}
The critical value $\mathcal{F}_{\mathrm{c}}$ of $\mathcal{F}$ for which
$\mathcal{D}=0.5$ at the end of the bifurcation line is given by
$\mathcal{F}_{\mathrm{c}}=0.5$.

Steady state bifurcation occurs when $\det J=0$, i.e. when%
\begin{equation}
0=3\mathcal{A}^{4}-4\mathcal{A}^{2}+1+\mathcal{D}^{2}\;.
\end{equation}
Substituting the solution, which is given by%
\begin{equation}
\mathcal{A}_{\pm}^{2}=\frac{2}{3}\pm\frac{1}{3}\sqrt{1-3\mathcal{D}^{2}}\;,
\end{equation}
into Eq. (\ref{F^2 C}) yields two branches%
\begin{align}
\mathcal{F}_{\pm}^{2}  &  =\left[  \left(  1-\frac{2}{3}\mp\frac{1}{3}%
\sqrt{1-3\mathcal{D}^{2}}\right)  ^{2}+\mathcal{D}^{2}\right] \nonumber\\
&  \times\left(  \frac{2}{3}\pm\frac{1}{3}\sqrt{1-3\mathcal{D}^{2}}\right)
\;.\nonumber\\
&  \label{F^2 pm}%
\end{align}
%

\begin{figure}
[ptb]
\begin{center}
\includegraphics[
height=3.3053in,
width=3.3347in
]%
{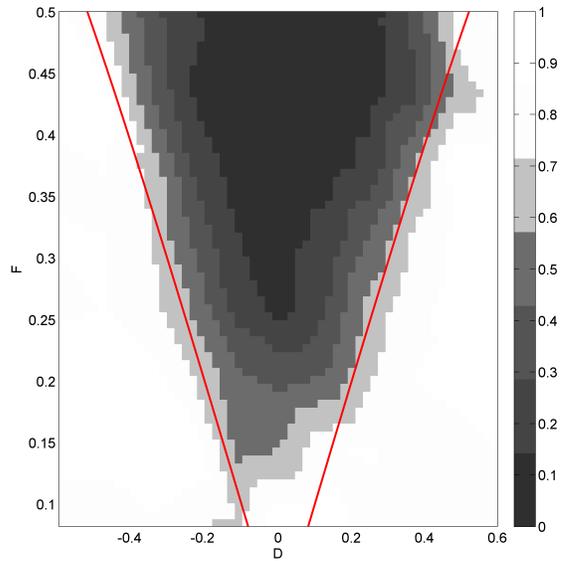}%
\caption{The normalized standard deviation $\sigma_{\phi}/\sigma_{\mathrm{u}}$
vs. $\mathcal{D}$ and $\mathcal{F}$. The solid line is the steady state
bifurcation line $\mathcal{F}_{-}\left(  \mathcal{D}\right)  $ [see Eq.
(\ref{F^2 pm})]. The device parameters are the same as those given in the
caption of Fig. \ref{Fig P_0 f_e}.}%
\label{Fig sync region}%
\end{center}
\end{figure}

Experimentally the region of synchronization is determined by measuring the
standard deviation of the phase of SEO, which is labeled as $\sigma_{\phi}$,
with varying values of the normalized detuning $\mathcal{D}$ and normalized
modulation amplitude $\mathcal{F}$. The measured normalized standard deviation
$\sigma_{\phi}/\sigma_{\mathrm{u}}$, where $\sigma_{\mathrm{u}}=1.8138$ is the
value corresponding to uniform distribution of the phase, is plotted in Fig.
\ref{Fig sync region}. In the region of phase locking, $\sigma_{\phi}%
/\sigma_{\mathrm{u}}\ll1$, whereas $\sigma_{\phi}/\sigma_{\mathrm{u}}\simeq1$
outside that region. The solid line is the steady state bifurcation line
$\mathcal{F}_{-}\left(  \mathcal{D}\right)  $ [see Eq. (\ref{F^2 pm})].
Theoretically, for $\mathcal{F}>\mathcal{F}_{\mathrm{c}}=0.5$ the region of
phase locking is expected to be determined by the Hopf bifurcation line given
by Eq. (\ref{F^2 Hopf}). However, this region is experimentally inaccessible
with the laser used in our experiment due to limited range of modulation amplitude.

\section{Summary}

In summary, synchronization in an on-fiber optomechanical cavity is
investigated. The relatively good agreement that is found between the
experimental results and the theoretical predictions validates the assumptions
and approximations that have been employed in the theoretical modeling. The
investigated device can be employed as a sensor operating in the region of
SEO. Future study will address the possibility of reducing phase noise by
inducing synchronization in order to enhance the sensor's performance.

\section{Acknowledgements}

This work was supported by the Israel Science Foundation, the bi-national
science foundation, the Security Research Foundation in the Technion, the
Israel Ministry of Science, the Russell Berrie Nanotechnology Institute and
MAGNET Metro 450 consortium.

\bibliographystyle{ieeepes}
\bibliography{acompat,Eyal_Bib}

\end{document}